%% file: main.tex
\documentclass[%
reprint,
amsmath,amssymb,superscriptaddress,
aps, prx
]{revtex4-2}

\usepackage{graphicx}
\usepackage{bbm}
\usepackage{xcolor}
\usepackage{hyperref}
\usepackage{physics}
\usepackage{dcolumn}

\begin{document}

\title{Adaptive variational algorithms for quantum Gibbs state preparation}

\author{Ada Warren}
    \affiliation{Department of Physics, Virginia Tech, Blacksburg, VA 24061, USA}
\author{Linghua Zhu}
    \affiliation{Department of Physics, Virginia Tech, Blacksburg, VA 24061, USA}
\author{Nicholas J. Mayhall}
    \affiliation{Department of Chemistry, Virginia Tech, Blacksburg, VA 24061, USA}
\author{Edwin Barnes}
    \affiliation{Department of Physics, Virginia Tech, Blacksburg, VA 24061, USA}
\author{Sophia E. Economou}
    \affiliation{Department of Physics, Virginia Tech, Blacksburg, VA 24061, USA}
    
\begin{abstract}
    The preparation of Gibbs thermal states is an important task in quantum computation with applications in quantum simulation, quantum optimization, and quantum machine learning. However, many algorithms for preparing Gibbs states rely on quantum subroutines which are difficult to implement on near-term hardware. Here, we address this by (i) introducing an objective function that, unlike the free energy, is easily measured, and (ii) using dynamically generated, problem-tailored ans\"atze. This allows for arbitrarily accurate Gibbs state preparation using low-depth circuits. To verify the effectiveness of our approach, we numerically demonstrate that our algorithm can prepare high-fidelity Gibbs states across a broad range of temperatures and for a variety of Hamiltonians.
\end{abstract}
    
\maketitle

\section{Introduction}

State preparation is a task of central importance in quantum computing. In particular, the preparation of finite-temperature thermal mixed states of a given Hamiltonian can be used for quantum simulation, quantum machine learning, and dynamics simulations of open systems~\cite{Kieferova2017PRA, Biamonte2017Nature, Somma2008PRL, poulin2009sampling, haug2020generalized}. This is challenging, and likely no efficient algorithm exists to solve the problem generally due to the complexity. Nevertheless, there are several proposed methods for sampling from thermal states of physically relevant Hamiltonians~\cite{terhal2000problem, poulin2009sampling, Temme2011Nature, Kastoryano2016CommMatPhys, Johri2017PRB, Brandao2019CommMatPhys}. Many of these methods, however, require use of costly quantum subroutines like quantum phase estimation or the estimation of von Neumann entropy, limiting prospects of thermal state preparation on near-term quantum hardware.

Variational quantum algorithms (VQAs) are a class of classical/quantum hybrid algorithms in which a quantum circuit is trained via classical optimization in order to reduce the value of some objective function which is evaluated on a quantum computer~\cite{mcclean2016theory, farhi2014quantum, hadfield2019quantum,cerezo2021variational}. These variational algorithms have been adapted to a variety of different quantum computing tasks, including quantum state preparation. The use of VQAs for thermal state preparation may allow for a reduction in the quantum resources required at the expense of performing a classical optimization. There has been interest recently in using VQAs to prepare Gibbs states~\cite{wu2019variational, wang2020variational,chowdhury2020variational, martyn2019product, Yuan2019arXiv, Motta2020NatPhys, zhu2020generation}. 

The Gibbs state for a data system $D$ represented by $N_D$ qubits with associated Hilbert space $\mathcal{H}_D$ and Hamiltonian $H$ at temperature $T$ is defined as the density operator $\rho _G = e^{-\beta H}/Z$, where $\beta = \frac{1}{k_B T}$ is the inverse temperature, $k_B$ is Boltzmann’s constant, and $Z = \Tr(e^{-\beta H})$ is the partition function. Because the target Gibbs state $\rho_G$ is a mixed state in general, deterministically preparing the thermal state requires entangling the data system $D$ with a purifying ancillary subsystem $A$ consisting of $N_A$ qubits with associated Hilbert space $\mathcal{H}_A$. The problem of Gibbs state preparation then resolves to preparing a pure state $\ket{\psi} \in \mathcal{H}_D \otimes \mathcal{H}_A$ such that $\Tr_A{\dyad{\psi}} = \rho_G$. This can be done variationally by preparing suitably-parameterized states $\ket{\theta} \in \mathcal{H}_D \otimes \mathcal{H}_A$ and classically optimizing the parameters $\theta$ to minimize an appropriate objective function $C(\ket{\theta})$.

Variational Gibbs state preparation typically relies on the well-known fact that the Gibbs state $\rho_G$ exactly minimizes the Gibbs free energy~\cite{wu2019variational}:
\begin{align}\label{eq:free_energy}
    F(\rho) &= E(\rho) - k_B T S(\rho) \\
        &= \Tr(\rho H) + \beta^{-1} \Tr(\rho \ln(\rho)) \nonumber.
\end{align}
However, the von Neumann entropy $S(\rho)$ and its gradients are difficult to measure on real quantum hardware~\cite{guo2021thermal,martyn2019product,wild2021quantum}, with the cost scaling exponentially with system size, particularly in the case of Gibbs states, as eigenvalues of the target state $\rho_G$ are exponentially suppressed~\cite{Acharya2020IEEE, Wang2022arXiv}. This makes the implementation of variational Gibbs state preparation too demanding for near-term quantum processors, especially at low temperatures. 

In this paper, we address this issue by introducing a new objective function which is minimized by the Gibbs state but, unlike the Gibbs free energy, does not require estimation of the von Neumann entropy. We then describe two different adaptive variational approaches for arbitrary-temperature Gibbs state preparation which make use of our new objective function. Rather than using pre-defined variational ans\"atze, we take inspiration from the recently-introduced Adaptive Derivative-Assembled Problem-Tailored Variational Quantum Algorithms (ADAPT-VQAs), which systematically build adaptive ans\"atze which are tailored to the given problem, providing better performance at lower circuit depth~\cite{grimsley2019adaptive, tang2021qubit, zhu2020qaoa}. We present numerical results demonstrating that both algorithms are capable of preparing high-fidelity Gibbs states across a range of temperatures for a few different Hamiltonians. Finally, we discuss evaluation of our new objective function on near-term quantum hardware.

\section{Objective function}

To avoid the challenge of estimating $S(\rho)$, we introduce a new objective function:
\begin{align} \label{eq:new_obj}
    C(\rho) &= -\Tr(\rho_G \rho) + \frac{1}{2}\Tr(\rho^2) \\
    &= -\frac{1}{Z}\Tr(e^{-\beta H} \rho) + \frac{1}{2}\Tr(\rho^2). \nonumber
\end{align}
Like the free energy, this new objective function is easily shown to be minimized by $\rho_G$. Unlike the free energy, however, measuring this objective function does not require estimating the von Neumann entropy. Instead, it relies only on the ability to measure an observable of the data system and the state purity, which can be measured using e.g. the SWAP test~\cite{Buhrman2001PRL, Ekert2002PRL, GarciaEscartin2013PRA, Cincio2018NewJPhys, wang2020variational}.

At first glance, this new objective function may appear to be of questionable value, as it seems to require prior knowledge of the operator $e^{-\beta H}$ and its trace, the partition function. As we will discuss later, however, we find that by Taylor expanding this operator and then truncating the resulting series, a suitable objective function for high-fidelity Gibbs states can still be obtained. This reduces the requirements for objective function estimation to measuring the purity and the first few powers of $H$.

In addition to evaluation of the objective function itself, evaluation of objective function gradients is of considerable importance for optimization algorithms. This is especially true for ADAPT-VQAs, which use information about objective function gradients to construct variational ans\"atze. In addition to permitting finite-difference methods of gradient estimation, in the special case that our parameterized state is of the form $\ket{\theta} = e^{i \theta G}\ket{\psi_0}$ where $G$ is an Hermitian operator with exactly two distinct eigenvalues $e_0$ and $e_1$, the exact gradient $\pdv{\theta} C(\rho(\theta))$, where $\rho(\theta) = \Tr_A \dyad{\theta}$, can be computed using a parameter-shift rule~\cite{crooks2019gradients, wang2020variational}. By introducing the auxiliary function
\begin{equation}
    \tilde{C}(\theta, \phi) = -\Tr(\rho_G \rho(\theta)) + \Tr(\rho(\theta) \rho(\phi)),
\end{equation}
which can be estimated with the same expectation value measurement and SWAP test circuits~\cite{Buhrman2001PRL, GarciaEscartin2013PRA} used to estimate $C(\rho(\theta))$, we can relate
\begin{equation} \label{eq:param-shift}
    \pdv{\theta} C(\rho(\theta)) = r\qty[\tilde{C}\qty(\theta + \frac{\pi}{4r}, \theta) - \tilde{C}\qty(\theta - \frac{\pi}{4r}, \theta)],
\end{equation}
where $r = \frac{1}{2}(e_1 - e_0)$.

\section{Prior work}

Ref.~\cite{wu2019variational} found success in preparing Gibbs states using a layered QAOA-like ansatz of the form
\begin{equation} \label{eq:qaoa-ansatz}
    \ket{\vec{\alpha}_n,\vec{\gamma}_n} = \prod_{k=1}^n e^{i \alpha_k H_{AD}} e^{i \gamma_k\qty(H_A + H_D) / 2} \ket{\psi_0},
\end{equation}
where $H_A + H_D = H \otimes \mathbbm{1} + \mathbbm{1} \otimes H$ is the problem Hamiltonian applied to both $A$ and $D$. $H_{AD} = \sum_{k=1}^{N_D}(X_{D_k}X_{A_k} + Y_{D_k}Y_{A_k} + Z_{D_k}Z_{A_k})$ is an operator whose ground state $\ket{\psi_0}$ exhibits maximal entanglement between $A$ and $D$, and is thus equivalent to the Gibbs state as $\beta \to 0$. Like the original QAOA algorithm, the algorithm of Ref.~\cite{wu2019variational} is assumed to work via a combination of the adiabatic theorem and trotterization, with the chosen operators allowing interpolation between $\ket{\psi_0}$---the ground state of $H_{AD}$---and the ground state of $H_A + H_D$, which is the $\beta \to \infty$ thermal state. This presumably allows preparation of thermal states at any temperature. However, based on this intuition, one would expect this approach to require an increasingly larger number of layers as one goes to lower temperatures, where entanglement is lowest. Essentially, the state preparation circuit is largely `undoing' all the entanglement that was purposely built in earlier by the circuit. This is borne out by the numerical results of Ref.~\cite{wu2019variational}, where the authors find that $N_D / 2$ layers are needed to realize high fidelity at low temperatures for a few different models, with faster convergence at higher temperatures. This is especially significant given that Gibbs state preparation is most challenging, and thus most interesting, at low temperatures. It may be possible to lessen the number of layers required by starting from a partially-entangled state, but as one would no longer be starting from a known thermal state, it is unclear that this ansatz structure would continue to see success. In addition, the ansatz structure and initial maximally-entangled state used in Ref.~\cite{wu2019variational} necessarily require $N_A = N_D$. While exact Gibbs state preparation at nonzero temperature is possible only when $N_A \geq N_D$, at low temperatures, many of the eigenvalues of $\rho_G$ become vanishingly small, implying that high-fidelity approximate Gibbs state preparation using smaller ancilla systems is possible~\cite{wang2020variational}. As we show below, if taken advantage of, this yields considerable quantum resource savings.

\section{ADAPT-VQE-Gibbs algorithm}

To limit the circuit depth of the state preparation circuit, we design a dynamically generated ansatz, which is by construction compact.
Inspired by the success of the ADAPT-VQE and qubit-ADAPT-VQE algorithms in finding effective, low-depth variational ansatze for molecular VQE problems~\cite{grimsley2019adaptive, tang2021qubit}, as well as the lack of restrictions the algorithm imposes on the initial reference state, we first present the ADAPT-VQE-Gibbs algorithm to prepare Gibbs states. After choosing an appropriate initial state $\ket{\psi_\text{ref}}$, the ansatz is grown iteratively, with only one additional operator $\tau_i = -\tau_i^\dagger$ added at each iteration. After the $n$-th iteration, the ansatz takes the form
\begin{equation}
	\ket{\vec{\theta}_n} = e^{\theta_n \tau_n} \ldots e^{\theta_2 \tau_2} e^{\theta_1 \tau_1} \ket{\psi_\text{ref}}.
\end{equation}
At the end of each iteration, an ordinary fixed-ansatz VQA optimization is employed to minimize the value of the objective function with respect to each of the classical parameters $\theta_i$. We call the optimized state after the $n$-th iteration $\ket{\vec{\theta}^*_n}$.
 
The operators $\tau_i$ are all chosen from a pre-defined operator pool and selected based on a gradient criterion, i.e. $\tau_{n}$ is chosen to maximize
\begin{equation}
    \abs{\eval{\pdv{\theta_n} C\qty(\Tr_A \dyad{\vec{\theta}_n})}_{\vec{\theta}_n = \mqty(0 & \vec{\theta}_n^*)}}.
\end{equation}
This process of iteratively adding pool operators and doing fixed-ansatz minimization is repeated until the norm of the vector of pool operator gradients falls below some pre-defined threshold $\epsilon$. 

To maximize the resource savings offered by utilizing the parameter-shift rule, as well as to reduce final circuit complexity, we choose a pool consisting of all 1- and 2-qubit Pauli strings acting on our combined data/ancilla system. Such a pool is clearly ``complete" in that it suffices to construct any unitary on the full data/ancilla system given enough layers, though more compact complete pools have been shown to exist for any number of qubits~\cite{tang2021qubit, shkolnikov2021avoiding}.

Compared to a pre-defined fixed-ansatz variational minimization, this ADAPT-VQA requires an increased number of measurements, requiring at least roughly twice as many measurements as computing the objective function itself, multiplied by the number of operators in the pool, for each iteration. This is in addition to all measurements required to do the fixed-ansatz minimization at the end of each iteration. Additionally, as the pool consists of only quite simple operators, each with little ability to change the prepared state overall, it is reasonable to assume that this will generate ans\"atze with considerably more classically-optimized parameters than the maximum of $N_D$ offered by Ref.~\cite{wu2019variational}. However, because it uses a gradient descent approach to operator selection, this algorithm allows for the construction of efficient, shallow-depth ans\"atze that still allow for high-fidelity Gibbs state preparation.

Care must be taken in the choice of initial state. When using ADAPT-VQE to find molecular ground states, one typically chooses some unentangled Fock basis state as the initial state. As we know that low-temperature Gibbs states are far from maximally entangled, this may seem to be an appropriate choice. Any state which leaves the data and ancilla system totally unentangled, however, maximizes the purity of the data system's reduced density matrix. For such states, then, gradients of the objective function come only from the term $\Tr(\rho(\theta) \rho_G)$, which is simply the expectation value of a Hermitian operator local to the data system. Thus, for our chosen pool and starting with an unentangled state, the largest operator gradient will always come from a local generator incapable of generating entanglement between the data and ancilla system, and thus the true nonzero-temperature Gibbs state will never be reached. For similar reasons, states for which the data and ancilla systems are maximally entangled are also typically unsuitable as initial states for ADAPT-VQE-Gibbs using this pool.

There are many options for preparing a suitable partially-entangled state. For our initial state, starting from $\ket{0}^{\otimes N_D + N_A}$, we apply one layer of random $y$-rotations to each qubit and then apply a CNOT gate between each ancilla-data qubit pair with the ancilla as the control:
\begin{equation}
    \ket{\psi_\text{ref}} = \prod_{i = 1}^{N_A}\prod_{j = 1}^{N_D} \text{CNOT}_{A_i D_j} \qty(\bigotimes_{i=1}^{N_A + N_D} e^{-i \alpha_i Y} \ket{0}), 
\end{equation}
where $\text{CNOT}_{A_i D_j}$ is a CNOT gate with qubit $i$ in $A$ as the control and qubit $j$ in $D$ as the target. These random values $\alpha_i \in \qty[0, 2\pi]$ are chosen at the beginning and serve only to provide a means of creating a suitable partially-entangled state. They are not optimized during subsequent steps. An unfortunate choice in these initial parameters may inadvertently lead to a situation as described above, where pool operator gradients end up vanishing despite not being near a thermal state, leading to premature termination of the algorithm. To guard against this, we repeat the algorithm several times, each with a difference choice for the initial parameters, and postselect the preparation which best minimizes the objective function.

To investigate the effectiveness of this approach, we simulate ADAPT-VQE-Gibbs for data systems consisting of $N_D = 4$ qubits for two Hamiltonians of interest: a 1-dimensional spin-$1/2$ Ising chain with periodic boundary conditions (i.e. $Z_{D,N_D + 1} = Z_{D,1}$):
\begin{equation} \label{eq:Ising_Ham}
    H_I = -\sum_{i = 1}^{N_D} Z_{D,i}Z_{D,i + 1},
\end{equation}
as well as a 1-dimensional spin-$1/2$ $XY$ chain with periodic boundary conditions:
\begin{equation}
    H_{XY} = -\sum_{i = 1}^{N_D} X_{D, i} X_{D, i + 1} + Y_{D, i} Y_{D, i + 1}.
\end{equation}
These two Hamiltonians have different spectra and eigenstates, leading to quite different thermal states. Additionally, as $H_I$ has degenerate ground states, preparing its Gibbs state requires nontrivial entanglement between $A$ and $D$ even in the limit $\beta \to \infty$.

\begin{figure*}
    \resizebox{\linewidth}{!}{\input{vqe-performance-tex}} 
    \caption{Plot of final Gibbs state fidelities and final ansatz CNOT counts across a range of temperatures for ADAPT-VQE-Gibbs with gradient threshold $\epsilon = 10^{-3}$ for \textbf{(a)} the periodic Ising Hamiltonian $H_I$ \textbf{(b)} the periodic XY Hamiltonian $H_{XY}$. The solid lines in the fidelity plots indicate the highest possible fidelity using a purifying system of $N_A$ qubits. CNOT counts for our ans\"atze include the $N_D N_A$ CNOTs used to prepare the reference state, as well as two CNOTs for every 2-qubit Pauli generator chosen to grow the ansatz. The dashed black lines in the CNOT count plots indicate the number of CNOTs required to execute the ansatz from Ref.~\cite{wu2019variational}, assuming the full $N_D / 2$ layers are required, as they are at low temperatures.}
    \label{fig:vqe-performance}
\end{figure*}
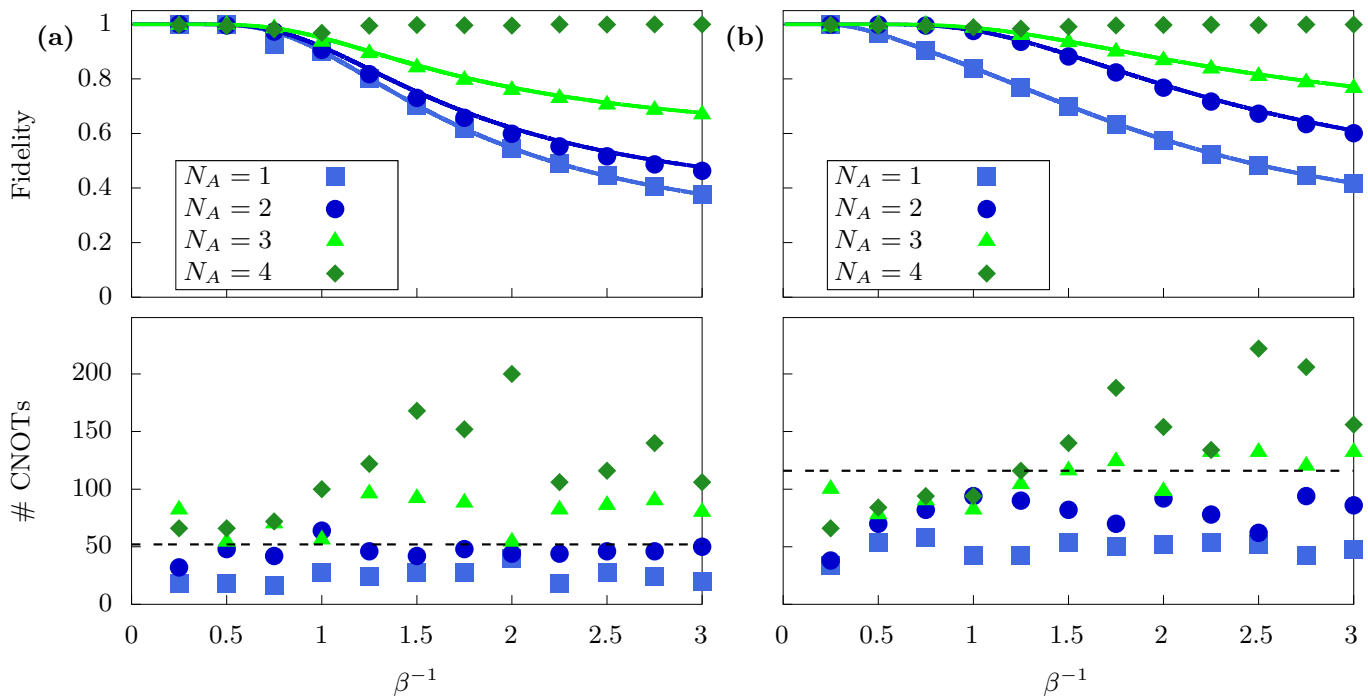

To investigate the effectiveness of this approach at restricted ancilla system size, we simulate for $N_A = \qty{1, 2, 3, 4}$. At each temperature and each ancilla system size, we run ADAPT-VQE-Gibbs in its entirety five times, then postselect the best-performing ansatz. To evaluate performance of the final ansatz, we calculate the fidelity of the final optimized state relative to the true Gibbs state $F = \Tr[\sqrt{\sqrt{\rho_G}\Tr_A \qty(\dyad{\vec{\theta}_{n_\text{max}}^*})\sqrt{\rho_G}}]^2$, where $n_\text{max}$ is the number of layers required to reach the gradient threshold $\epsilon$. We show the results of these simulations in Fig.~\ref{fig:vqe-performance}. We find that, for $N_A = N_D$, ADAPT-VQE-Gibbs is able to consistently achieve $>99\%$ fidelity. For reduced $N_A$, ADAPT-VQE-Gibbs is still able to reach the largest possible fidelity given the restricted space of accessible purified data system states. While this gives quite poor fidelity at high temperatures, very high fidelities can still be reached with reduced $N_A$ at low temperatures, allowing for a considerable reduction in quantum resources. Additionally, at low temperatures for $H_{XY}$, we find that our algorithm reaches high fidelity using fewer CNOTs than the ansatz used by Ref.~\cite{wu2019variational} for a system of this size. This is true even for $N_A = N_D$, where perfect Gibbs state preparation is guaranteed to be possible. For $H_I$, Ref.~\cite{wu2019variational}permits reducing the entangling Hamiltonian to $H_{AD} = \sum_{k=1}^{N_D} X_{D_k}X_{A_k}$, allowing for a sizable reduction in the number of CNOTs used. In this case, our algorithm actually uses more two-qubit operations at $N_A = N_D$. At reduced $N_A$, however---particularly for $N_A = 1$ or $2$---we still see a reduction in the number of CNOTs used over Ref.~\cite{wu2019variational} without sacrificing fidelity.

\section{ADAPT-QAOA-Gibbs algorithm}

At higher temperatures, ADAPT-VQE-Gibbs requires an increasing number of resources, particularly at $N_A = N_D$. These are also the conditions under which the ansatz of Ref.~\cite{wu2019variational} converges fastest, often requiring fewer than $N_D/2$ layers to reach high fidelity. Thus, at intermediate to large temperatures, ADAPT-VQE-Gibbs may offer little to no advantage. It may still be possible, however, to realize an improvement over the ansatz of Ref.~\cite{wu2019variational} at these temperatures using an ADAPT-VQA which still utilizes a QAOA-like structure. 

Inspired by the successes of ADAPT-QAOA in finding efficient ans\"atze for combinatorial QAOA problems~\cite{zhu2020qaoa}, we now introduce ADAPT-QAOA-Gibbs---a modification of the ADAPT-QAOA algorithm that can be used to prepare Gibbs states. To briefly review: in ADAPT-QAOA, the layered structure of the QAOA ansatz (each layer consisting of evolution under the cost Hamiltonian followed by evolution under a mixer operator) is retained, but the use of a single, fixed mixer operator is abandoned. Instead, the ansatz is grown iteratively, layer by layer, with each layer's mixer chosen adaptively from a pre-defined pool based on gradients of the cost Hamiltonian. This yields a problem-tailored ansatz which outperforms the original QAOA ansatz. Similarly, in ADAPT-QAOA-Gibbs, we retain the QAOA-like layered structure employed by Ref.~\cite{wu2019variational} (Eq. (\ref{eq:qaoa-ansatz})), with each layer consisting of evolution under the problem Hamiltonian followed by evolution under an entangling Hamiltonian. Instead of using the fixed entangling Hamiltonian $H_{AD}$, however, we will similarly select entangling operators $\tau_i = -\tau_i^\dagger$ from a pre-defined pool, adaptively generating a problem-tailored ansatz, which at the $n$-th iteration takes the form
\begin{equation}
    \ket{\vec{\alpha}_n,\vec{\gamma}_n} = \prod_{k=1}^n e^{\alpha_k \tau_k} e^{i \gamma_k\qty(H_A + H_D) / 2} \ket{\psi_0}.
\end{equation}
For ADAPT-QAOA-Gibbs, we always choose $N_A = N_D$, and start in the maximally-entangled reference state $\ket{\psi_0}$.
After each iteration, an ordinary fixed-ansatz VQA is employed to minimize the objective function with respect to each of the classical parameters $\alpha_i$ and $\gamma_i$. We call the optimized state after the $n$-th iteration $\ket{\vec{\alpha}_n^*,\vec{\gamma}_n^*}$.

For ADAPT-QAOA-Gibbs, we choose an operator pool consisting of all 2-qubit Pauli strings which act nontrivially on both $A$ and $D$, as well the operator $i H_{AD}$. Exactly which operator is used for a given layer is determined by measuring gradients of the objective function and choosing the pool operator which yields the largest gradient, i.e., $\tau_n$ is chosen to maximize
\begin{equation}
    \abs{\eval{\pdv{\alpha_n}C\qty(\Tr_A\dyad{\vec{\alpha}_n,\vec{\gamma}_n})}_{\vec{\alpha}_n = \mqty( 0 & \vec{\alpha}_{n-1}^* )}^{\vec{\gamma}_n = \mqty(\gamma_0 & \vec{\gamma}_{n-1}^*)}},
\end{equation}
where $\gamma_0 \in [0, \pi/2]$ is chosen randomly at the beginning of the algorithm and remains constant throughout. To guard against unfortunate choices of $\gamma_0$, we run the algorithm several times and postselect the ansatz which best minimizes the objective function. All of these gradients can be measured via the parameter-shift rule, Eq. (\ref{eq:param-shift}). This can be done directly for all of the Pauli string operators. The remaining operator $H_{AD}$ is a sum of mutually-commuting Pauli strings, and its gradient can be decomposed into the sum of gradients of those operators~\cite{crooks2019gradients}. The presence of the operator $H_{AD}$ in the entangler pool, while not necessary to reach convergence, ensures that ADAPT-QAOA-Gibbs generates ans\"atze which differ from those of Ref.~\cite{wu2019variational} only in the event that another operator in the pool is found to have a larger gradient. It is thus reasonable to assume that the final generated ans\"atze will converge to the Gibbs state at least as quickly as those of Ref.~\cite{wu2019variational}. As every other operator in the pool other than $H_{AD}$ can be implemented using fewer 2-qubit interactions, this implies that our ans\"atze can be implemented on quantum hardware using equal or fewer resources.

\begin{figure}
    \resizebox{\linewidth}{!}{\input{qaoa-performance-tex}}
    \caption{\textbf{(a)} Layer-by-layer performance of ADAPT-QAOA-Gibbs for a periodic Ising chain with size $N_D = 6$. Gibbs state fidelity exceeding $99\%$ is achieved by the third layer or sooner for all temperatures. \textbf{(b)} Number of CNOTs required to reach $\geq 99\%$ at each temperature. Dashed line indicates number of CNOTs required by the ansatz used in Ref.~\cite{wu2019variational} to reach $\geq 99\%$.} 
    \label{fig:adapt-qaoa-temp}
\end{figure}
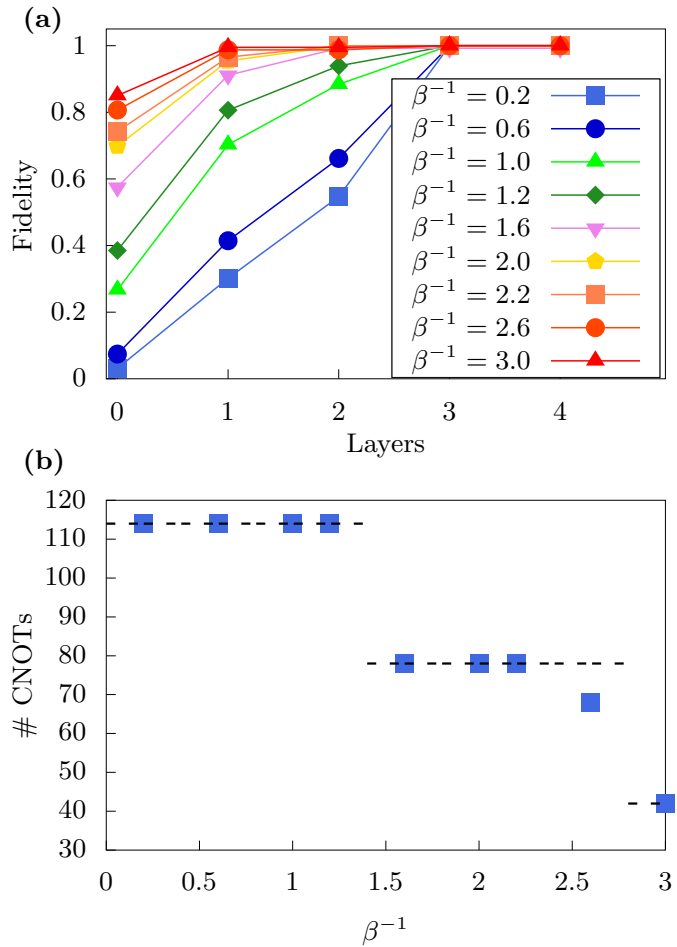

To investigate the effectiveness of this approach, we simulate ADAPT-QAOA-Gibbs at a variety of temperatures for a data system consisting of $N_D = 6$ qubits under a 1-dimensional spin-$1/2$ Ising chain Hamiltonian with periodic boundary conditions~\ref{eq:Ising_Ham}. For each temperature, we execute four iterative steps of the ADAPT-QAOA-Gibbs algorithm for eight different initializations of $\gamma_0$, and postselect the best-performing ansatz. To evaluate performance, after each layer, we compute the fidelity of the optimized state $\ket{\vec{\alpha}_n^*,\vec{\gamma}_n^*}$ relative to the true Gibbs state. The results of these simulations are shown in Fig.~\ref{fig:adapt-qaoa-temp}. We find that we achieve $>99\%$ fidelity across a range of temperatures using three layers or fewer, reaching convergence more quickly at larger temperatures. This is consistent with the $N_D/2$ layers cited by Ref.~\cite{wu2019variational}. We also find that, with the exception of $\beta^{-1} = 2.6$, ADAPT-QAOA-Gibbs simply reproduces the ansatz of Ref.~\cite{wu2019variational}, providing no circuit advantage. At $\beta^{-1} = 2.6$, however, we find that savings of $10$ CNOT gates is achieved. We expect that larger savings should be possible for larger systems, which tend to require more layers.

\section{Measuring the objective function}

Measurement of the objective function $C(\rho)$ requires estimation of the expectation value $\Tr(\rho_G \rho)$, and the state purity $\Tr(\rho^2)$. The state purity is easily measured using e.g. the SWAP test, which requires preparation of two identical copies of the prepared state, as well as a circuit consisting of $N_A$ cSWAP gates~\cite{Buhrman2001PRL}. Estimation of the expectation value $\Tr(\rho_G \rho)$ presents more of a challenge, as the Hermitian operator $e^{-\beta H}$ is not typically known exactly \textit{a priori}. Assuming, however, that measurements of $H^n$ are accessible for small integers $n$, we can approximate this operator by Taylor expanding and then truncating to finite order $m$:
\begin{equation}
	e^{-\beta H} \approx \sum_{n=0}^m \frac{1}{n!}\qty(-\beta)^n H^n.
\end{equation}
This truncated operator divided by its trace can be used to define a truncated objective function that can be used for Gibbs state preparation.

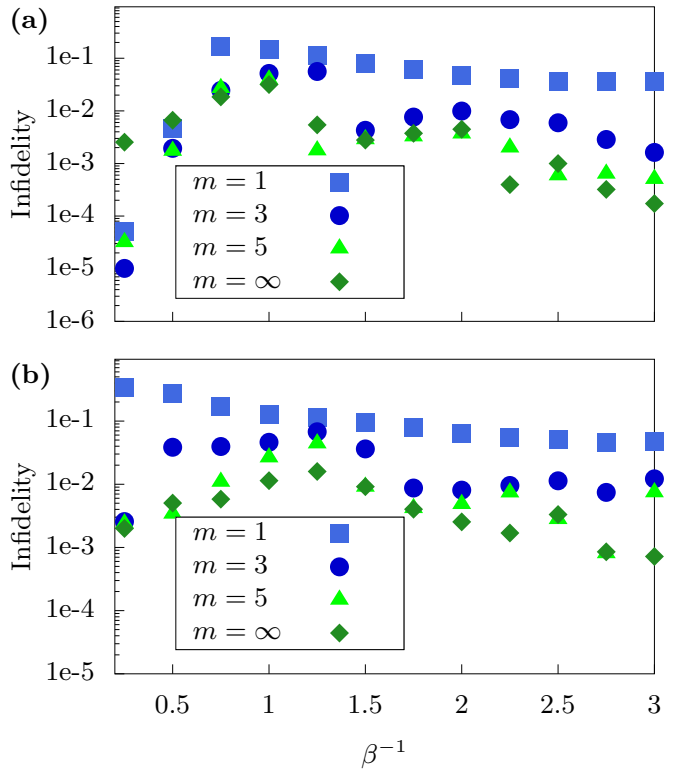
\begin{figure}
    \resizebox{\linewidth}{!}{\input{vqe-order-comparison-tex}} 
    \caption{Logarithmic plots of ADAPT-VQE-Gibbs final Gibbs state infidelity with a gradient threshold of $\epsilon = 10^{-3}$ and $N_A = N_D = 4$ for various temperatures and truncation orders for \textbf{(a)} the periodic Ising Hamiltonian $H_I$ \textbf{(b)} the periodic XY Hamiltonian $H_{XY}$. Each data point represents postselected best-performing ans\"atze among five different random initializations.}
    \label{fig:truncation_infidelity}
\end{figure}

This truncation shifts the minimum of the objective function away from the desired exact Gibbs state, limiting the fidelity of the final prepared state. We explore the effect of this truncation for our adaptive Gibbs state preparation algorithms in Fig.~\ref{fig:truncation_infidelity}. While very low-order truncations yield unfavorable results, performance at $m = 5$ is found to be on par with the non-truncated infinite series despite the considerable error in approximating the operator $\rho_G$ in the objective function. This suggests that even moderately low-order truncation is still suitable for high-fidelity Gibbs state preparation. While not shown, performance of ADAPT-QAOA-Gibbs is similar.

Assuming the number of operators in $H$ scales as $\mathcal{O}(N_D)$, measuring $H^k$ requires estimating $\mathcal{O}(N_D^k)$ different observables. The exact cost of estimating the truncated objective function for Gibbs state preparation then depends on the truncation order $m$ required to achieve high-fidelity Gibbs states.

\section{Conclusions}

To conclude, we have introduced a new objective function suitable for variational Gibbs state preparation which does not rely upon measurement of the von Neumann entropy---a task which is known to be quite difficult. We also introduced two new ADAPT-VQAs capable of using our objective function to adaptively generate effective ans\"atze for the preparation of high-fidelity Gibbs states. Our first algorithm was able to produce high-fidelity Gibbs states with significant resource savings compared to previous methods at low temperature, while our second algorithm was able to squeeze additional performance out of known methods at intermediate temperatures. This suggests that ADAPT-VQAs utilizing novel objective functions could provide a path to near-term thermal state preparation.

\section*{Acknowledgments}

This work was supported by the Department of Energy. S.E.E. acnowledges the DOE Office of Science, National Quantum Information Science Research Centers, Co-design Center for Quantum Advantage (C2QA), contract number DE-SC0012704. E.B. and N.J.M. acknowledge award No. DE-SC0019199. 

\bibliography{bibliography.bib}

\end{document}

%% file: vqe-performance-tex.tex
\begingroup
  \makeatletter
  \providecommand\color[2][]{%
    \GenericError{(gnuplot) \space\space\space\@spaces}{%
      Package color not loaded in conjunction with
      terminal option `colourtext'%
    }{See the gnuplot documentation for explanation.%
    }{Either use 'blacktext' in gnuplot or load the package
      color.sty in LaTeX.}%
    \renewcommand\color[2][]{}%
  }%
  \providecommand\includegraphics[2][]{%
    \GenericError{(gnuplot) \space\space\space\@spaces}{%
      Package graphicx or graphics not loaded%
    }{See the gnuplot documentation for explanation.%
    }{The gnuplot epslatex terminal needs graphicx.sty or graphics.sty.}%
    \renewcommand\includegraphics[2][]{}%
  }%
  \providecommand\rotatebox[2]{#2}%
  \@ifundefined{ifGPcolor}{%
    \newif\ifGPcolor
    \GPcolortrue
  }{}%
  \@ifundefined{ifGPblacktext}{%
    \newif\ifGPblacktext
    \GPblacktexttrue
  }{}%
  \let\gplgaddtomacro\g@addto@macro
  \gdef\gplbacktext{}%
  \gdef\gplfronttext{}%
  \makeatother
  \ifGPblacktext
    \def\colorrgb#1{}%
    \def\colorgray#1{}%
  \else
    \ifGPcolor
      \def\colorrgb#1{\color[rgb]{#1}}%
      \def\colorgray#1{\color[gray]{#1}}%
      \expandafter\def\csname LTw\endcsname{\color{white}}%
      \expandafter\def\csname LTb\endcsname{\color{black}}%
      \expandafter\def\csname LTa\endcsname{\color{black}}%
      \expandafter\def\csname LT0\endcsname{\color[rgb]{1,0,0}}%
      \expandafter\def\csname LT1\endcsname{\color[rgb]{0,1,0}}%
      \expandafter\def\csname LT2\endcsname{\color[rgb]{0,0,1}}%
      \expandafter\def\csname LT3\endcsname{\color[rgb]{1,0,1}}%
      \expandafter\def\csname LT4\endcsname{\color[rgb]{0,1,1}}%
      \expandafter\def\csname LT5\endcsname{\color[rgb]{1,1,0}}%
      \expandafter\def\csname LT6\endcsname{\color[rgb]{0,0,0}}%
      \expandafter\def\csname LT7\endcsname{\color[rgb]{1,0.3,0}}%
      \expandafter\def\csname LT8\endcsname{\color[rgb]{0.5,0.5,0.5}}%
    \else
      \def\colorrgb#1{\color{black}}%
      \def\colorgray#1{\color[gray]{#1}}%
      \expandafter\def\csname LTw\endcsname{\color{white}}%
      \expandafter\def\csname LTb\endcsname{\color{black}}%
      \expandafter\def\csname LTa\endcsname{\color{black}}%
      \expandafter\def\csname LT0\endcsname{\color{black}}%
      \expandafter\def\csname LT1\endcsname{\color{black}}%
      \expandafter\def\csname LT2\endcsname{\color{black}}%
      \expandafter\def\csname LT3\endcsname{\color{black}}%
      \expandafter\def\csname LT4\endcsname{\color{black}}%
      \expandafter\def\csname LT5\endcsname{\color{black}}%
      \expandafter\def\csname LT6\endcsname{\color{black}}%
      \expandafter\def\csname LT7\endcsname{\color{black}}%
      \expandafter\def\csname LT8\endcsname{\color{black}}%
    \fi
  \fi
    \setlength{\unitlength}{0.0500bp}%
    \ifx\gptboxheight\undefined%
      \newlength{\gptboxheight}%
      \newlength{\gptboxwidth}%
      \newsavebox{\gptboxtext}%
    \fi%
    \setlength{\fboxrule}{0.5pt}%
    \setlength{\fboxsep}{1pt}%
    \definecolor{tbcol}{rgb}{1,1,1}%
\begin{picture}(9216.00,4608.00)%
    \gplgaddtomacro\gplbacktext{%
      \csname LTb\endcsname
      \put(559,2557){\makebox(0,0)[r]{\strut{}0}}%
      \put(559,2930){\makebox(0,0)[r]{\strut{}0.2}}%
      \put(559,3303){\makebox(0,0)[r]{\strut{}0.4}}%
      \put(559,3675){\makebox(0,0)[r]{\strut{}0.6}}%
      \put(559,4048){\makebox(0,0)[r]{\strut{}0.8}}%
      \put(559,4421){\makebox(0,0)[r]{\strut{}1}}%
      \put(691,2337){\makebox(0,0){\strut{} }}%
      \put(1340,2337){\makebox(0,0){\strut{} }}%
      \put(1989,2337){\makebox(0,0){\strut{} }}%
      \put(2638,2337){\makebox(0,0){\strut{} }}%
      \put(3286,2337){\makebox(0,0){\strut{} }}%
      \put(3935,2337){\makebox(0,0){\strut{} }}%
      \put(4584,2337){\makebox(0,0){\strut{} }}%
      \put(42,4328){\makebox(0,0)[l]{\strut{}\textbf{(a)}}}%
    }%
    \gplgaddtomacro\gplfronttext{%
      \csname LTb\endcsname
      \put(-46,3535){\rotatebox{-270}{\makebox(0,0){\strut{}Fidelity}}}%
      \put(2637,2051){\makebox(0,0){\strut{}}}%
      \csname LTb\endcsname
      \put(1653,3379){\makebox(0,0)[r]{\strut{}$N_A = 1$}}%
      \csname LTb\endcsname
      \put(1653,3159){\makebox(0,0)[r]{\strut{}$N_A = 2$}}%
      \csname LTb\endcsname
      \put(1653,2939){\makebox(0,0)[r]{\strut{}$N_A = 3$}}%
      \csname LTb\endcsname
      \put(1653,2719){\makebox(0,0)[r]{\strut{}$N_A = 4$}}%
      \csname LTb\endcsname
      \put(2637,45917){\makebox(0,0){\strut{}}}%
    }%
    \gplgaddtomacro\gplbacktext{%
      \csname LTb\endcsname
      \put(5005,2557){\makebox(0,0)[r]{\strut{} }}%
      \put(5005,2930){\makebox(0,0)[r]{\strut{} }}%
      \put(5005,3303){\makebox(0,0)[r]{\strut{} }}%
      \put(5005,3675){\makebox(0,0)[r]{\strut{} }}%
      \put(5005,4048){\makebox(0,0)[r]{\strut{} }}%
      \put(5005,4421){\makebox(0,0)[r]{\strut{} }}%
      \put(5137,2337){\makebox(0,0){\strut{} }}%
      \put(5786,2337){\makebox(0,0){\strut{} }}%
      \put(6435,2337){\makebox(0,0){\strut{} }}%
      \put(7084,2337){\makebox(0,0){\strut{} }}%
      \put(7732,2337){\makebox(0,0){\strut{} }}%
      \put(8381,2337){\makebox(0,0){\strut{} }}%
      \put(9030,2337){\makebox(0,0){\strut{} }}%
      \put(4748,4328){\makebox(0,0)[l]{\strut{}\textbf{(b)}}}%
    }%
    \gplgaddtomacro\gplfronttext{%
      \csname LTb\endcsname
      \put(4774,3535){\rotatebox{-270}{\makebox(0,0){\strut{}}}}%
      \put(7083,2051){\makebox(0,0){\strut{}}}%
      \csname LTb\endcsname
      \put(6099,3379){\makebox(0,0)[r]{\strut{}$N_A = 1$}}%
      \csname LTb\endcsname
      \put(6099,3159){\makebox(0,0)[r]{\strut{}$N_A = 2$}}%
      \csname LTb\endcsname
      \put(6099,2939){\makebox(0,0)[r]{\strut{}$N_A = 3$}}%
      \csname LTb\endcsname
      \put(6099,2719){\makebox(0,0)[r]{\strut{}$N_A = 4$}}%
      \csname LTb\endcsname
      \put(7083,50431){\makebox(0,0){\strut{}}}%
    }%
    \gplgaddtomacro\gplbacktext{%
      \csname LTb\endcsname
      \put(559,460){\makebox(0,0)[r]{\strut{}0}}%
      \put(559,853){\makebox(0,0)[r]{\strut{}50}}%
      \put(559,1246){\makebox(0,0)[r]{\strut{}100}}%
      \put(559,1640){\makebox(0,0)[r]{\strut{}150}}%
      \put(559,2033){\makebox(0,0)[r]{\strut{}200}}%
      \put(691,240){\makebox(0,0){\strut{}0}}%
      \put(1340,240){\makebox(0,0){\strut{}0.5}}%
      \put(1989,240){\makebox(0,0){\strut{}1}}%
      \put(2638,240){\makebox(0,0){\strut{}1.5}}%
      \put(3286,240){\makebox(0,0){\strut{}2}}%
      \put(3935,240){\makebox(0,0){\strut{}2.5}}%
      \put(4584,240){\makebox(0,0){\strut{}3}}%
    }%
    \gplgaddtomacro\gplfronttext{%
      \csname LTb\endcsname
      \put(-46,1439){\rotatebox{-270}{\makebox(0,0){\strut{}\# CNOTs}}}%
      \put(2637,-90){\makebox(0,0){\strut{}$\beta^{-1}$}}%
      \csname LTb\endcsname
      \put(2637,52849){\makebox(0,0){\strut{}}}%
    }%
    \gplgaddtomacro\gplbacktext{%
      \csname LTb\endcsname
      \put(5005,460){\makebox(0,0)[r]{\strut{} }}%
      \put(5005,853){\makebox(0,0)[r]{\strut{} }}%
      \put(5005,1246){\makebox(0,0)[r]{\strut{} }}%
      \put(5005,1640){\makebox(0,0)[r]{\strut{} }}%
      \put(5005,2033){\makebox(0,0)[r]{\strut{} }}%
      \put(5137,240){\makebox(0,0){\strut{}0}}%
      \put(5786,240){\makebox(0,0){\strut{}0.5}}%
      \put(6435,240){\makebox(0,0){\strut{}1}}%
      \put(7084,240){\makebox(0,0){\strut{}1.5}}%
      \put(7732,240){\makebox(0,0){\strut{}2}}%
      \put(8381,240){\makebox(0,0){\strut{}2.5}}%
      \put(9030,240){\makebox(0,0){\strut{}3}}%
    }%
    \gplgaddtomacro\gplfronttext{%
      \csname LTb\endcsname
      \put(4774,1439){\rotatebox{-270}{\makebox(0,0){\strut{}}}}%
      \put(7083,-90){\makebox(0,0){\strut{}$\beta^{-1}$}}%
      \csname LTb\endcsname
      \put(7083,55267){\makebox(0,0){\strut{}}}%
    }%
    \gplbacktext
    \put(0,0){\includegraphics[width={460.80bp},height={230.40bp}]{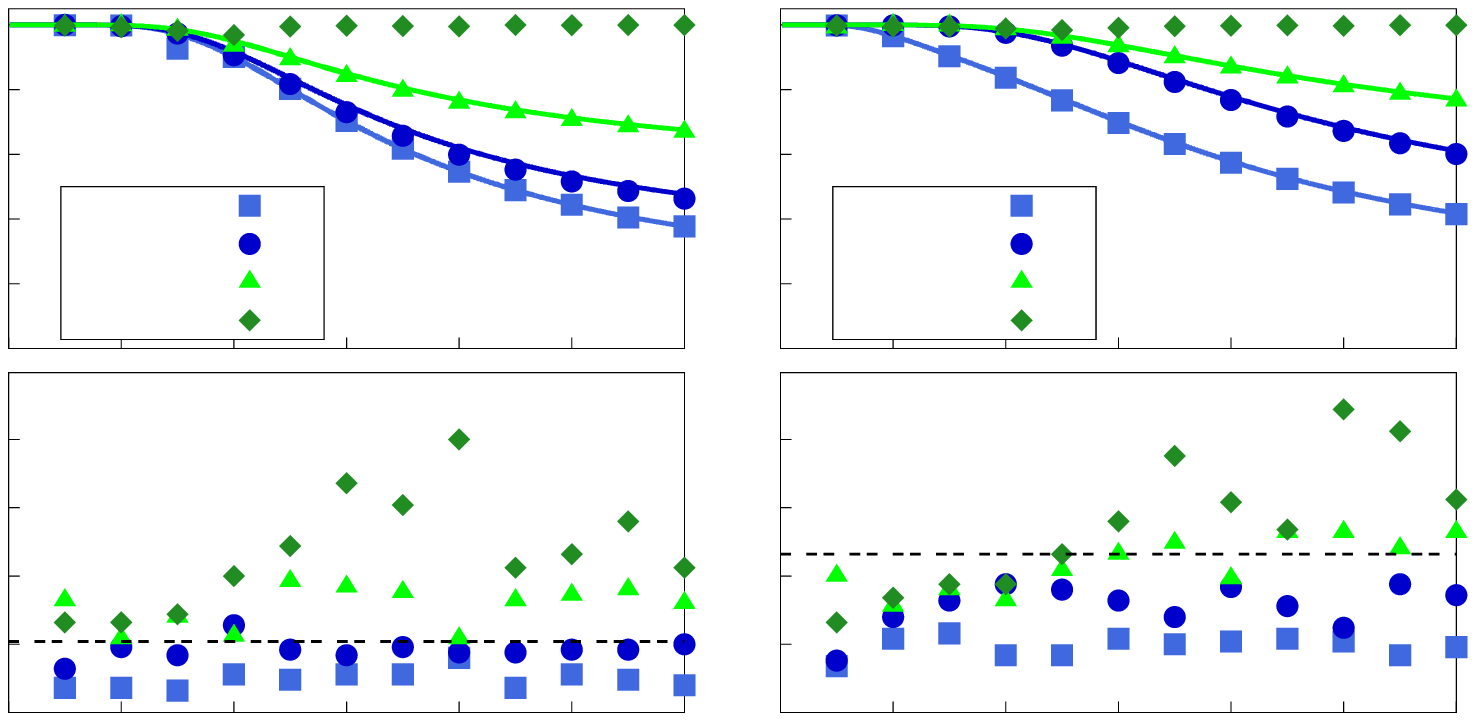}}%
    \gplfronttext
  \end{picture}%
\endgroup

%% file: qaoa-performance-tex.tex
\begingroup
  \makeatletter
  \providecommand\color[2][]{%
    \GenericError{(gnuplot) \space\space\space\@spaces}{%
      Package color not loaded in conjunction with
      terminal option `colourtext'%
    }{See the gnuplot documentation for explanation.%
    }{Either use 'blacktext' in gnuplot or load the package
      color.sty in LaTeX.}%
    \renewcommand\color[2][]{}%
  }%
  \providecommand\includegraphics[2][]{%
    \GenericError{(gnuplot) \space\space\space\@spaces}{%
      Package graphicx or graphics not loaded%
    }{See the gnuplot documentation for explanation.%
    }{The gnuplot epslatex terminal needs graphicx.sty or graphics.sty.}%
    \renewcommand\includegraphics[2][]{}%
  }%
  \providecommand\rotatebox[2]{#2}%
  \@ifundefined{ifGPcolor}{%
    \newif\ifGPcolor
    \GPcolortrue
  }{}%
  \@ifundefined{ifGPblacktext}{%
    \newif\ifGPblacktext
    \GPblacktexttrue
  }{}%
  \let\gplgaddtomacro\g@addto@macro
  \gdef\gplbacktext{}%
  \gdef\gplfronttext{}%
  \makeatother
  \ifGPblacktext
    \def\colorrgb#1{}%
    \def\colorgray#1{}%
  \else
    \ifGPcolor
      \def\colorrgb#1{\color[rgb]{#1}}%
      \def\colorgray#1{\color[gray]{#1}}%
      \expandafter\def\csname LTw\endcsname{\color{white}}%
      \expandafter\def\csname LTb\endcsname{\color{black}}%
      \expandafter\def\csname LTa\endcsname{\color{black}}%
      \expandafter\def\csname LT0\endcsname{\color[rgb]{1,0,0}}%
      \expandafter\def\csname LT1\endcsname{\color[rgb]{0,1,0}}%
      \expandafter\def\csname LT2\endcsname{\color[rgb]{0,0,1}}%
      \expandafter\def\csname LT3\endcsname{\color[rgb]{1,0,1}}%
      \expandafter\def\csname LT4\endcsname{\color[rgb]{0,1,1}}%
      \expandafter\def\csname LT5\endcsname{\color[rgb]{1,1,0}}%
      \expandafter\def\csname LT6\endcsname{\color[rgb]{0,0,0}}%
      \expandafter\def\csname LT7\endcsname{\color[rgb]{1,0.3,0}}%
      \expandafter\def\csname LT8\endcsname{\color[rgb]{0.5,0.5,0.5}}%
    \else
      \def\colorrgb#1{\color{black}}%
      \def\colorgray#1{\color[gray]{#1}}%
      \expandafter\def\csname LTw\endcsname{\color{white}}%
      \expandafter\def\csname LTb\endcsname{\color{black}}%
      \expandafter\def\csname LTa\endcsname{\color{black}}%
      \expandafter\def\csname LT0\endcsname{\color{black}}%
      \expandafter\def\csname LT1\endcsname{\color{black}}%
      \expandafter\def\csname LT2\endcsname{\color{black}}%
      \expandafter\def\csname LT3\endcsname{\color{black}}%
      \expandafter\def\csname LT4\endcsname{\color{black}}%
      \expandafter\def\csname LT5\endcsname{\color{black}}%
      \expandafter\def\csname LT6\endcsname{\color{black}}%
      \expandafter\def\csname LT7\endcsname{\color{black}}%
      \expandafter\def\csname LT8\endcsname{\color{black}}%
    \fi
  \fi
    \setlength{\unitlength}{0.0500bp}%
    \ifx\gptboxheight\undefined%
      \newlength{\gptboxheight}%
      \newlength{\gptboxwidth}%
      \newsavebox{\gptboxtext}%
    \fi%
    \setlength{\fboxrule}{0.5pt}%
    \setlength{\fboxsep}{1pt}%
    \definecolor{tbcol}{rgb}{1,1,1}%
\begin{picture}(4320.00,6192.00)%
    \gplgaddtomacro\gplbacktext{%
      \csname LTb\endcsname
      \put(386,3746){\makebox(0,0)[r]{\strut{}0}}%
      \put(386,4188){\makebox(0,0)[r]{\strut{}0.2}}%
      \put(386,4630){\makebox(0,0)[r]{\strut{}0.4}}%
      \put(386,5072){\makebox(0,0)[r]{\strut{}0.6}}%
      \put(386,5514){\makebox(0,0)[r]{\strut{}0.8}}%
      \put(386,5956){\makebox(0,0)[r]{\strut{}1}}%
      \put(592,3526){\makebox(0,0){\strut{}0}}%
      \put(1327,3526){\makebox(0,0){\strut{}1}}%
      \put(2062,3526){\makebox(0,0){\strut{}2}}%
      \put(2798,3526){\makebox(0,0){\strut{}3}}%
      \put(3533,3526){\makebox(0,0){\strut{}4}}%
      \put(-37,6133){\makebox(0,0)[l]{\strut{}\textbf{(a)}}}%
    }%
    \gplgaddtomacro\gplfronttext{%
      \csname LTb\endcsname
      \put(-21,4906){\rotatebox{-270}{\makebox(0,0){\strut{}Fidelity}}}%
      \put(2375,3306){\makebox(0,0){\strut{}Layers}}%
      \csname LTb\endcsname
      \put(3340,5625){\makebox(0,0)[r]{\strut{}$\hspace{1mm}\beta^{-1} = 0.2$}}%
      \csname LTb\endcsname
      \put(3340,5405){\makebox(0,0)[r]{\strut{}$\hspace{1mm}\beta^{-1} = 0.6$}}%
      \csname LTb\endcsname
      \put(3340,5185){\makebox(0,0)[r]{\strut{}$\hspace{1mm}\beta^{-1} = 1.0$}}%
      \csname LTb\endcsname
      \put(3340,4965){\makebox(0,0)[r]{\strut{}$\hspace{1mm}\beta^{-1} = 1.2$}}%
      \csname LTb\endcsname
      \put(3340,4745){\makebox(0,0)[r]{\strut{}$\hspace{1mm}\beta^{-1} = 1.6$}}%
      \csname LTb\endcsname
      \put(3340,4525){\makebox(0,0)[r]{\strut{}$\hspace{1mm}\beta^{-1} = 2.0$}}%
      \csname LTb\endcsname
      \put(3340,4305){\makebox(0,0)[r]{\strut{}$\hspace{1mm}\beta^{-1} = 2.2$}}%
      \csname LTb\endcsname
      \put(3340,4085){\makebox(0,0)[r]{\strut{}$\hspace{1mm}\beta^{-1} = 2.6$}}%
      \csname LTb\endcsname
      \put(3340,3865){\makebox(0,0)[r]{\strut{}$\hspace{1mm}\beta^{-1} = 3.0$}}%
      \csname LTb\endcsname
      \put(2375,61334){\makebox(0,0){\strut{}}}%
    }%
    \gplgaddtomacro\gplbacktext{%
      \csname LTb\endcsname
      \put(386,619){\makebox(0,0)[r]{\strut{}30}}%
      \put(386,877){\makebox(0,0)[r]{\strut{}40}}%
      \put(386,1135){\makebox(0,0)[r]{\strut{}50}}%
      \put(386,1393){\makebox(0,0)[r]{\strut{}60}}%
      \put(386,1651){\makebox(0,0)[r]{\strut{}70}}%
      \put(386,1908){\makebox(0,0)[r]{\strut{}80}}%
      \put(386,2166){\makebox(0,0)[r]{\strut{}90}}%
      \put(386,2424){\makebox(0,0)[r]{\strut{}100}}%
      \put(386,2682){\makebox(0,0)[r]{\strut{}110}}%
      \put(386,2940){\makebox(0,0)[r]{\strut{}120}}%
      \put(518,399){\makebox(0,0){\strut{}0}}%
      \put(1137,399){\makebox(0,0){\strut{}0.5}}%
      \put(1756,399){\makebox(0,0){\strut{}1}}%
      \put(2375,399){\makebox(0,0){\strut{}1.5}}%
      \put(2994,399){\makebox(0,0){\strut{}2}}%
      \put(3613,399){\makebox(0,0){\strut{}2.5}}%
      \put(4232,399){\makebox(0,0){\strut{}3}}%
      \put(-37,3198){\makebox(0,0)[l]{\strut{}\textbf{(b)}}}%
    }%
    \gplgaddtomacro\gplfronttext{%
      \csname LTb\endcsname
      \put(-21,1779){\rotatebox{-270}{\makebox(0,0){\strut{}\# CNOTs}}}%
      \put(2375,69){\makebox(0,0){\strut{}$\beta^{-1}$}}%
      \csname LTb\endcsname
      \put(2375,64274){\makebox(0,0){\strut{}}}%
    }%
    \gplbacktext
    \put(0,0){\includegraphics[width={216.00bp},height={309.60bp}]{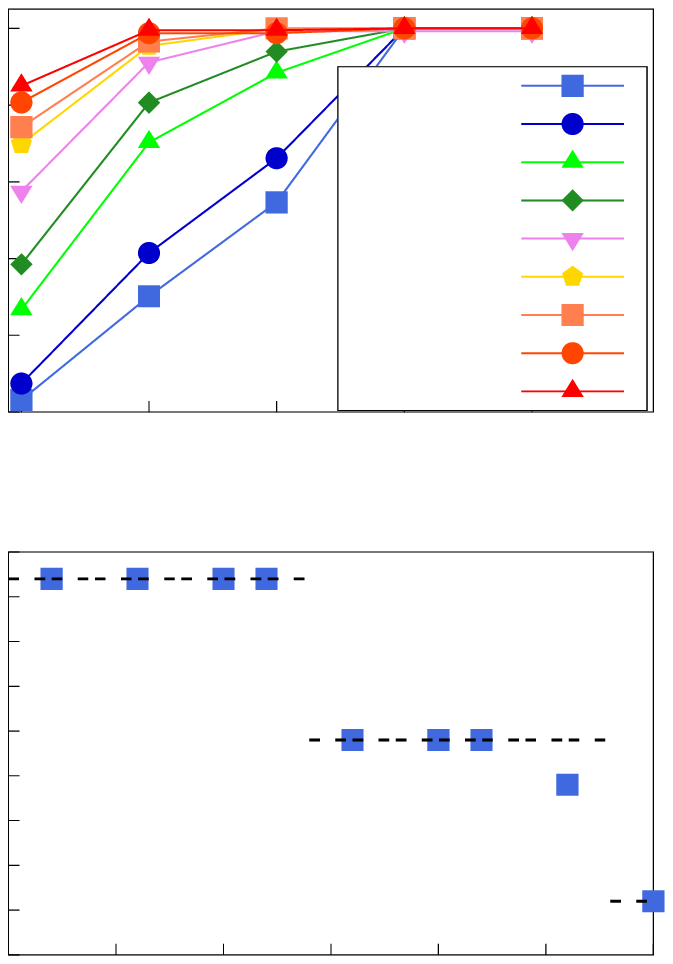}}%
    \gplfronttext
  \end{picture}%
\endgroup

%% file: vqe-order-comparison-tex.tex
\begingroup
  \makeatletter
  \providecommand\color[2][]{%
    \GenericError{(gnuplot) \space\space\space\@spaces}{%
      Package color not loaded in conjunction with
      terminal option `colourtext'%
    }{See the gnuplot documentation for explanation.%
    }{Either use 'blacktext' in gnuplot or load the package
      color.sty in LaTeX.}%
    \renewcommand\color[2][]{}%
  }%
  \providecommand\includegraphics[2][]{%
    \GenericError{(gnuplot) \space\space\space\@spaces}{%
      Package graphicx or graphics not loaded%
    }{See the gnuplot documentation for explanation.%
    }{The gnuplot epslatex terminal needs graphicx.sty or graphics.sty.}%
    \renewcommand\includegraphics[2][]{}%
  }%
  \providecommand\rotatebox[2]{#2}%
  \@ifundefined{ifGPcolor}{%
    \newif\ifGPcolor
    \GPcolortrue
  }{}%
  \@ifundefined{ifGPblacktext}{%
    \newif\ifGPblacktext
    \GPblacktexttrue
  }{}%
  \let\gplgaddtomacro\g@addto@macro
  \gdef\gplbacktext{}%
  \gdef\gplfronttext{}%
  \makeatother
  \ifGPblacktext
    \def\colorrgb#1{}%
    \def\colorgray#1{}%
  \else
    \ifGPcolor
      \def\colorrgb#1{\color[rgb]{#1}}%
      \def\colorgray#1{\color[gray]{#1}}%
      \expandafter\def\csname LTw\endcsname{\color{white}}%
      \expandafter\def\csname LTb\endcsname{\color{black}}%
      \expandafter\def\csname LTa\endcsname{\color{black}}%
      \expandafter\def\csname LT0\endcsname{\color[rgb]{1,0,0}}%
      \expandafter\def\csname LT1\endcsname{\color[rgb]{0,1,0}}%
      \expandafter\def\csname LT2\endcsname{\color[rgb]{0,0,1}}%
      \expandafter\def\csname LT3\endcsname{\color[rgb]{1,0,1}}%
      \expandafter\def\csname LT4\endcsname{\color[rgb]{0,1,1}}%
      \expandafter\def\csname LT5\endcsname{\color[rgb]{1,1,0}}%
      \expandafter\def\csname LT6\endcsname{\color[rgb]{0,0,0}}%
      \expandafter\def\csname LT7\endcsname{\color[rgb]{1,0.3,0}}%
      \expandafter\def\csname LT8\endcsname{\color[rgb]{0.5,0.5,0.5}}%
    \else
      \def\colorrgb#1{\color{black}}%
      \def\colorgray#1{\color[gray]{#1}}%
      \expandafter\def\csname LTw\endcsname{\color{white}}%
      \expandafter\def\csname LTb\endcsname{\color{black}}%
      \expandafter\def\csname LTa\endcsname{\color{black}}%
      \expandafter\def\csname LT0\endcsname{\color{black}}%
      \expandafter\def\csname LT1\endcsname{\color{black}}%
      \expandafter\def\csname LT2\endcsname{\color{black}}%
      \expandafter\def\csname LT3\endcsname{\color{black}}%
      \expandafter\def\csname LT4\endcsname{\color{black}}%
      \expandafter\def\csname LT5\endcsname{\color{black}}%
      \expandafter\def\csname LT6\endcsname{\color{black}}%
      \expandafter\def\csname LT7\endcsname{\color{black}}%
      \expandafter\def\csname LT8\endcsname{\color{black}}%
    \fi
  \fi
    \setlength{\unitlength}{0.0500bp}%
    \ifx\gptboxheight\undefined%
      \newlength{\gptboxheight}%
      \newlength{\gptboxwidth}%
      \newsavebox{\gptboxtext}%
    \fi%
    \setlength{\fboxrule}{0.5pt}%
    \setlength{\fboxsep}{1pt}%
    \definecolor{tbcol}{rgb}{1,1,1}%
\begin{picture}(4320.00,5040.00)%
    \gplgaddtomacro\gplbacktext{%
      \csname LTb\endcsname
      \put(516,2847){\makebox(0,0)[r]{\strut{}1e-6}}%
      \put(516,3197){\makebox(0,0)[r]{\strut{}1e-5}}%
      \put(516,3547){\makebox(0,0)[r]{\strut{}1e-4}}%
      \put(516,3896){\makebox(0,0)[r]{\strut{}1e-3}}%
      \put(516,4246){\makebox(0,0)[r]{\strut{}1e-2}}%
      \put(516,4596){\makebox(0,0)[r]{\strut{}1e-1}}%
      \put(1032,2627){\makebox(0,0){\strut{} }}%
      \put(1672,2627){\makebox(0,0){\strut{} }}%
      \put(2312,2627){\makebox(0,0){\strut{} }}%
      \put(2952,2627){\makebox(0,0){\strut{} }}%
      \put(3592,2627){\makebox(0,0){\strut{} }}%
      \put(4232,2627){\makebox(0,0){\strut{} }}%
      \put(-54,4840){\makebox(0,0)[l]{\strut{}\textbf{(a)}}}%
    }%
    \gplgaddtomacro\gplfronttext{%
      \csname LTb\endcsname
      \put(43,3892){\rotatebox{-270}{\makebox(0,0){\strut{}Infidelity}}}%
      \put(2440,2341){\makebox(0,0){\strut{}}}%
      \csname LTb\endcsname
      \put(1053,3770){\makebox(0,0)[l]{\strut{}$\hspace{2mm} m = 1$}}%
      \csname LTb\endcsname
      \put(1053,3550){\makebox(0,0)[l]{\strut{}$\hspace{2mm} m = 3$}}%
      \csname LTb\endcsname
      \put(1053,3330){\makebox(0,0)[l]{\strut{}$\hspace{2mm} m = 5$}}%
      \csname LTb\endcsname
      \put(1053,3110){\makebox(0,0)[l]{\strut{}$\hspace{2mm} m = \infty$}}%
      \csname LTb\endcsname
      \put(2440,38808){\makebox(0,0){\strut{}}}%
    }%
    \gplgaddtomacro\gplbacktext{%
      \csname LTb\endcsname
      \put(516,504){\makebox(0,0)[r]{\strut{}1e-5}}%
      \put(516,924){\makebox(0,0)[r]{\strut{}1e-4}}%
      \put(516,1344){\makebox(0,0)[r]{\strut{}1e-3}}%
      \put(516,1764){\makebox(0,0)[r]{\strut{}1e-2}}%
      \put(516,2184){\makebox(0,0)[r]{\strut{}1e-1}}%
      \put(1032,284){\makebox(0,0){\strut{}0.5}}%
      \put(1672,284){\makebox(0,0){\strut{}1}}%
      \put(2312,284){\makebox(0,0){\strut{}1.5}}%
      \put(2952,284){\makebox(0,0){\strut{}2}}%
      \put(3592,284){\makebox(0,0){\strut{}2.5}}%
      \put(4232,284){\makebox(0,0){\strut{}3}}%
      \put(-54,2478){\makebox(0,0)[l]{\strut{}\textbf{(b)}}}%
    }%
    \gplgaddtomacro\gplfronttext{%
      \csname LTb\endcsname
      \put(43,1549){\rotatebox{-270}{\makebox(0,0){\strut{}Infidelity}}}%
      \put(2440,-46){\makebox(0,0){\strut{}$\beta^{-1}$}}%
      \csname LTb\endcsname
      \put(1053,1435){\makebox(0,0)[l]{\strut{}$\hspace{2mm} m = 1$}}%
      \csname LTb\endcsname
      \put(1053,1215){\makebox(0,0)[l]{\strut{}$\hspace{2mm} m = 3$}}%
      \csname LTb\endcsname
      \put(1053,995){\makebox(0,0)[l]{\strut{}$\hspace{2mm} m = 5$}}%
      \csname LTb\endcsname
      \put(1053,775){\makebox(0,0)[l]{\strut{}$\hspace{2mm} m = \infty$}}%
      \csname LTb\endcsname
      \put(2440,41403){\makebox(0,0){\strut{}}}%
    }%
    \gplbacktext
    \put(0,0){\includegraphics[width={216.00bp},height={252.00bp}]{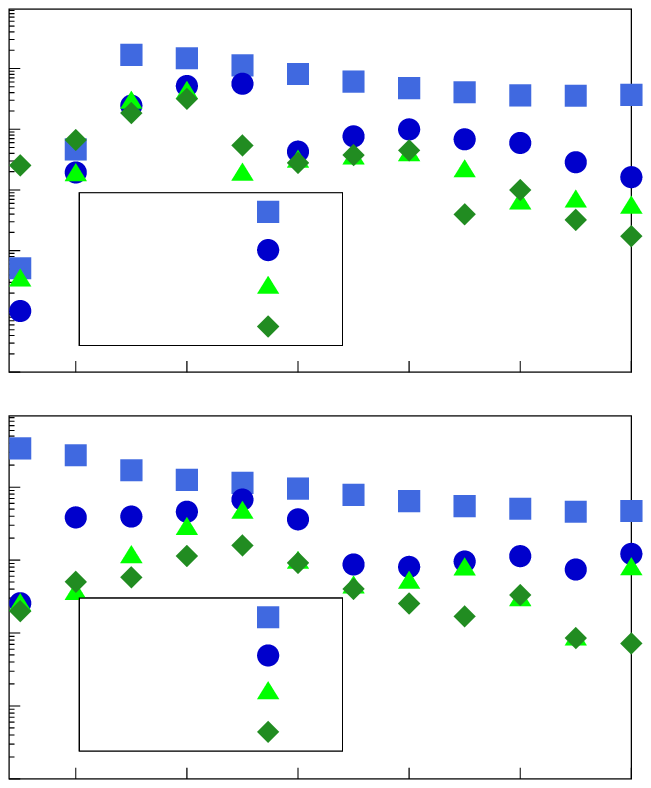}}%
    \gplfronttext
  \end{picture}%
\endgroup